\documentclass[prd,twocolumn,showpacs,letterpaper,nofootinbib]{revtex4}
\usepackage{graphics}
\usepackage{epsfig} 
\usepackage{amsmath}

\def\persec{{\rm s}^{-1}}

\def\erg{{\rm erg}}
\def\eV{{\rm eV}}
\def\keV{{\rm keV}}
\def\MeV{{\rm MeV}}
\def\GeV{{\rm GeV}}
\def\TeV{{\rm TeV}}
\def\cm{{\rm cm}}
\def\sr{{\rm sr}}

\begin{document}

\title{Particle decays during the cosmic dark ages}

\author{Xuelei Chen}
\email{xuelei@kitp.ucsb.edu.}
\affiliation{The Kavli Institute for Theoretical Physics, 
UCSB, Santa Barbara, CA~~93106, USA.}
\author{Marc Kamionkowski}
\email{kamion@tapir.caltech.edu.}
\affiliation{Mail Code 130-33, California Institute of Technology, 
Pasadena, CA~~91125, USA.}

\date{{October 15, 2003}}

\preprint{}

\begin{abstract}
We consider particle decays during the cosmic dark
ages with two aims:  (1) to explain the high
optical depth reported by WMAP, and (2) to provide new
constraints to the parameter space for
decaying particles.  We delineate the decay channels
in which most of the decay energy ionizes and heats the 
IGM gas (and thus affects the CMB), 
and those in which most of the
energy is carried away---e.g. photons with energies $100~ \keV
\lesssim E \lesssim 1 ~\TeV$---and thus appears as a
contribution to diffuse x-ray and gamma-ray backgrounds.
The new constraints to the decay-particle parameters from the
CMB power spectrum thus complement those from the cosmic X-ray
and $\gamma$-ray backgrounds.  Although decaying particles can
indeed produce an optical depth consistent with that reported by
WMAP, in so doing they produce new fluctuations in the CMB
temperature/polarization power spectra.  For decay lifetimes
less than the age of the Universe, the induced power spectra
generally violate current constraints, while the power spectra
are usually consistent if the lifetime is longer than the age of the Universe.
\end{abstract}


\maketitle

\section{Introduction}
A large correlation between the temperature and E-type 
polarization at large angular scale (low $l$) was recently
observed by the Wilkinson Microwave Anisotropy Probe (WMAP)
\cite{WMAP}. This is a unique signature of re-scattering of cosmic 
microwave background (CMB)
photons at redshifts relatively low compared with that of the
last-scattering surface at $z \approx 1100$ \cite{TEreion}.
The required optical depth of $\tau_e \sim 0.17$ can be achieved
if reionization occurs at a redshift of $z_{re} \sim 20$.
Although there are theoretical uncertainties, such a
reionization redshift is difficult to reconcile with the
star-formation history expected in the $\Lambda$CDM model
\cite{starreion}, which generally favors a reionization redshift
of 7--12 \cite{reion-pre-wmap}.  Furthermore, the thermal
history of the intergalactic medium (IGM) contains further
evidence for late completion of reionization \cite{late-reion}.
This potential conflict between the evidence for early and late
reionization might be partially resolved in the
double-reionization model, where an early generation of massive,
metal-free stars were formed and partly ionized the Universe
\cite{reion-twice}. Nevertheless, even in this model, it is not
easy to achieve such a high optical depth \cite{WMAP}. 

In light of this, it is worthwhile to consider possible
alternatives. For example, it has been suggested that a high optical
depth might be achieved if primordial density fluctuations are
non-Gaussian \cite{C03}. Here we consider another option.  While
stellar photons must have
contributed to reionization, it remains possible that 
other energy sources also contribute.
Decay of an unstable particle, for example, provides such 
an alternative energy source. In this scenario, a decaying particle, possibly
part of the dark matter, releases energy during its decay, which
contributes to the ionization of the IGM. 
Another widely discussed 
possibility is the radiative decay of an active neutrino, which might
play a role in a number of astrophysical phenomena
\cite{sciama,historical,dodelson}.  
Although the parameters of the original model are
now excluded by observations \cite{decay_obs}, there are still
other regions of decaying-neutrino parameter space, and there is
no lack of other particle-physics candidates; e.g., unstable
supersymmetric particles \cite{susydecay}, cryptons \cite{ELN90},
R-parity violating gravitinos \cite{BMV91}, moduli dark matter
\cite{AHKY98}, superheavy dark-matter particles \cite{CKR99,DN02}, 
axinos \cite{KK02}, sterile neutrinos \cite{HH03}, weakly interacting
massive particles (WIMPs) decaying to superweakly interacting 
massive particles(SWIMPs) \cite{FRT03}, and quintessinos \cite{BLZ03}. 
Recently, Hansen and Haiman \cite{HH03} suggested
sterile-neutrino decay as a source of reionization. In addition
to decaying particles, evaporation of 
primordial black holes \cite{HF02} and decay of 
topological defects such as cosmic strings and monopoles are also
possible source of extra energy input. The decay of an unstable 
particle may also help explain the presence of dwarf spheroidal 
galaxies in the Local Group, resolve the cuspy halo problem 
in $\Lambda$CDM models \cite{C01a,C01b,S03}, and serve as a possible
source of the ultra high energy cosmic rays \cite{UHECR}.

From a cosmological perspective, it is particularly interesting to
consider the rich variety of ionization histories 
offered by the particle-decay scenario. In these scenarios, the
Universe is not necessarily fully ionized; instead, particle
decay may ionize only a small fraction of the gas. If the process
lasts for a large range of redshifts, it may still contribute a large
fraction of the measured free-electron optical depth. The presence of 
a not significantly damped first acoustic peak in the CMB anisotropy
spectrum suggests that particle decay should not significantly 
delay the recombination process at $z \sim 1100$ \cite{PSH00,BMS03},
but is it possible that the Universe become partially ionized during
the cosmic ``dark ages'' at redshifts of ten to a few hundred?  What is
the observational signature of such an ionization history?  Can
this scenario be distinguished from late reionization by CMB
observations?  Particle decay may also produce energetic
photons; can observation of cosmic $\gamma$-ray backgrounds
place constraints on this scenario?

In this paper we consider these questions. Since at low redshift
stars and quasars emit ionizing photons, and since at the
epoch of recombination there is no significant increase of
entropy, we shall focus mostly on particle decays in the
redshift range between 1000 and 20.  Such particles produce an
optical depth $\tau\sim0.17$ by partially reionizing the
Universe at redshifts much higher than the value, $z\sim20$,
required if the Universe becomes fully ionized by early star
formation.  We calculate the CMB temperature/polarization power
spectra induced by this alternative ionization history and show
that it can be distinguished from the full-reionization scenario
with the same $\tau$.  In some regions of the decay-particle
parameter space, the induced power spectra conflict with
those observed already, but there are other regions where decaying
particles can provide the required optical depth and maintain
consistency with the measured power spectra.

While investigating decaying particles as contributors to cosmic
reionization, it becomes clear that new CMB constraints to the
ionization history provide new constraints to the parameter space
for decaying particles.  To a first approximation,
the energy injected by particles that decay with
lifetimes between the ages of the Universe at recombination and
today either gets absorbed by the IGM, or it appears in diffuse
radiation backgrounds \cite{marc}.  In the latter case, observed radiation
backgrounds have traditionally been used to constrain the
parameter space that consists of the decay-particle lifetime and
density as well as the energy of the decay products.  As we
detail below, new CMB constraints to the ionization history can
now provide complementary new constraints to the regions of
parameter space where the decay energy goes to heating and
ionizing the IGM.

This paper is organized as follows: In the next Section,  
we discuss how energy is dissipated
for various decay channels and what fraction of energy is
eventually used for ionization, heating
the gas, or carried away by escaping photons and neutrino.  We
also discuss how this energy is deposited as a function of
redshift. In Section III,  we describe how to calculate the
ionization  history and CMB anisotropy with extra energy input
from decaying particles, and we discuss how the result depends
on the property of the particle.  In Section IV, we obtain
constraints to the decay-particle parameter space from the CMB
and diffuse backgrounds. We summarize our results in
Section V.\footnote{While this paper was being prepared, two
preprints \cite{kks03,P03} on similar questions appeared.  Our
results agree with theirs where our calculations overlap.}

\section{Decaying channels and energy dissipation}

Depending on the nature of the decaying particle, the decay products
may include gauge bosons, charged leptons, neutrinos, quarks, or other
more exotic particles. These particles may then subsequently
decay further into other particles, or they may interact
with particles in the IGM.  With sufficient energy, a shower of
particles is created. In the end, stable, weakly-interacting
particles like neutrinos escape, while other particles lose 
a significant part of their energy during the interaction 
with the primordial gas or cosmic microwave background. Some of
this energy can go into ionizing the IGM, and the efficiency of
converting the decay energy to ionization energy is process
dependent.  Here we review some of the more generic features; in
particular, we consider the efficiency of converting the rest
mass of the decaying particle to ionization energy, $\chi_i
\equiv E_i/M_X c^2$, where $M_x$ is the mass of the decaying
particle.

\begin{figure}
\epsfig{file=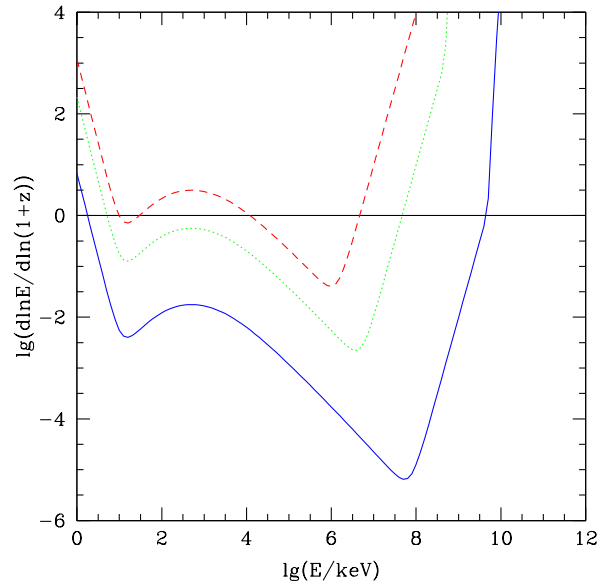,width=0.45\textwidth}
\caption{\label{gloss} Energy loss rate by photons per Hubble time. 
The solid (blue), 
dotted (green), dashed (red) line are for redshift $z+1=10, 100$, and $316$, 
respectively. We took $x_e=0.01$ here, but the results are insensitive to 
$x_e$ for $E > 1~\keV$.}
\end{figure}

\subsection{Photons} 
In this paper we are mostly interested in the ``dark ages'',
$10< z < 1000$, where most of the gas is neutral. 
Photons with energy smaller than 13.6 eV cannot ionize
hydrogen atoms in the ground state, but if there is a large
presence of hydrogen atoms in excited states, e.g. at the end of
the recombination era, $z \sim 1000$, photons with energy
$E<13.6$ eV may contribute to the ionization. When most of the
atoms fall to the ground state, photons with energy less than
13.6 eV will escape.

Ultraviolet and soft X-ray photons with energy $13.6 ~\eV-1 ~\keV$ 
have large photoionization cross sections and are largely
absorbed locally. For photons with energy $E>40 ~\eV$, neutral
helium absorption dominates the absorption, and photoelectrons
are produced in the process. The photoelectron carries the
remaining energy of the photon.  From here on the energy
dissipation process for the initial photon is the same as that
for an initial energetic electron.

The absorption processes of hard X-ray and $\gamma$-ray 
photons at cosmological distances were discussed in
Ref.~\cite{ZS89}. The processes in which photons
can be absorbed or lose energy include (i) photoionization of
atoms, (ii) Compton scattering on electrons,
(iii) production of pairs on atoms, 
(iv) production of pairs on free electrons and free nuclei, 
(v) scattering with background photons, and
(vi) pair production on background photons. In some of these,
such as Compton scattering, a photon loses only a small fraction
of its energy in a single scattering event, while in others it
can lose a significant part of its energy.

In Fig.~\ref{gloss}, we plot the total energy-loss rate,
\begin{equation}
-\frac{d \ln E}{d \ln (1+z)} = \frac{\Delta E}{E}\frac{n(z) \sigma(E) c}{H(z)},
\end{equation}
as a function of energy for redshifts,
$1+z=10, 100$, and $316$. We assume $\Delta E/E \sim 1$ except
for the Compton-scattering process.  Here, $H(z)$ is the
expansion rate of the Universe at redshift $z$,
$n(z)$ is the density of the target particle---i.e., neutral
hydrogen or helium for (i) or (iii), free electrons for (ii) or
(iv), and CMB photons for (v) or (vi)---and $\sigma$ 
is the corresponding cross section.  At the high energy ($E \ll 13.6 eV$)
where Compton scattering becomes important, a photon is 
not able to distinguish whether an electron is free
or bound \footnote{This was pointed out to us by Professor R.
Sunyaev, see e.g. \cite{SC96} for more detailed discussion.}, 
so we assume a free-electron fraction of 0.01 for processes
(i), (iii), (iv), but treat all electrons as free for the Compton 
processes. Our result is not sensitive to the ionization fraction as
long as the Universe is mostly neutral. We should also point out that
in this discussion we have also neglected any other photon background,
e.g. the infrared photon background which might be produced at low redshift.

As we can see from the Figure, high-energy photons (above 1 TeV at 
$z \sim 10$, 1 GeV at $z \sim 300$) can scatter with CMB photons or 
produce pairs, and they are largely absorbed locally, producing either
an X-ray photon which has a larger absorption cross section,
or an electron-positron pair with high energy. 

For photons with energy around 1 GeV (and around 1 keV at low redshift), 
the Universe is optically thin in the redshift range which we
are interested in, photons lose most of their energy by
redshifting, only a small fraction of the total energy is
transfered to gas by scattering, and the scattering occurs over
a wide range of redshifts. The details of the energy
distribution depend on the injection energy and redshift of the
photon.  The formation of an electromagnetic shower and the
resulting spectrum was investigated for both low redshift
\cite{Z88,ZS90,DN02} and the early Universe
\cite{KM95,PSB95}. The spectrum of the shower is of the form
$E^{-\alpha}$ below the threshold energy, with $0< \alpha < 2$
\cite{G92}.

\begin{figure}
\epsfig{file=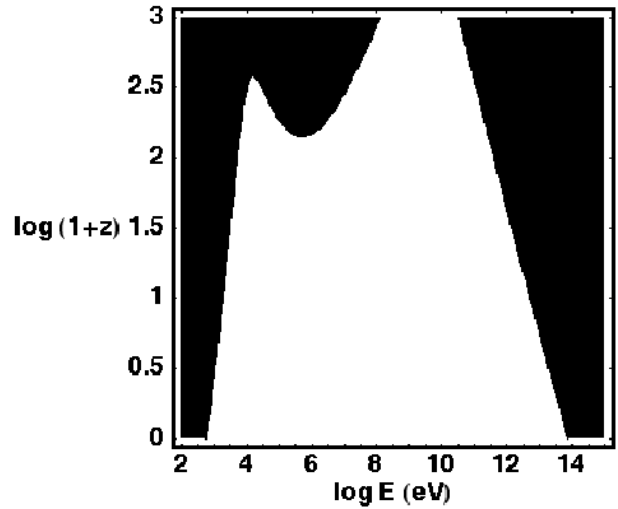,width=0.45\textwidth}
\caption{\label{gcontour} Transparency window for photons.  The
dark regions are those in which $d\log E/d\log(1+z)>1$; i.e.,
those in which most of the photon energy gets absorbed by the
IGM in a Hubble time.  In the clear regions, the Universe is
transparent to photons.}
\end{figure}

Fig. \ref{gcontour} shows the transparency window.  In the dark
regions, $d\log E/d\log(1+z)>1$ (i.e., most of the photon energy
goes to the IGM), and in the white regions $d\log
E/d\log(1+z)<1$ (i.e., the Universe is transparent to these
photons).  The bump in the transparency window
at ($\log E, \log(1+z)$)=(5.7, 2.16) is 
due to the Compton scattering. 
If the photon is injected in the transparency window,
and remains in the transparency window as its energy redshifts
(once a photon is injected it travels down and to the left on
this plot), then it will free stream and appear in diffuse
radiation backgrounds with energies $\sim$keV--10 TeV;
otherwise, it will not appear as a diffuse background, but will
heat and ionize the IGM.  Also keep in mind that the time
interval $dt \propto dz (1+z)^{-5/2}$.  Thus, if a particle
decays with lifetime longer than the age of the Universe, the
relevant redshift for determining whether it appears in diffuse
backgrounds is $z=0$.  In Section IV below, we will give
constraints to the decay-particle parameter space under two
extreme assumptions:  (1) that all the photon energy goes into
the IGM, and (2) that the photons free stream and appear as
diffuse backgrounds.  Fig. \ref{gcontour} must then be consulted
to determine which limit applies for a particular decay-photon
energy and lifetime.

What we need to know for the ionization history is what fraction
of the energy is converted to ionization energy, and how it is distributed over
redshift. In the optically-thick case---i.e. for photons outside
the transparency window---we can assume that the energy is
instantly deposited.  The energy deposition rate (in units of
$\erg~ \cm^{-3}~ \persec$) is 
\begin{equation}
\label{eq:Qinstant}
Q(z) = \chi_i(z) n_X(z) M_X c^2 \Gamma_X,
\end{equation}
where $M_X$ is the mass of the decaying particle, and $\Gamma_X$
the decay rate. If the lifetime of the particle is much longer
than the age of the Universe and $\chi$ is constant, then $Q(z)
\propto (1+z)^3$.

In the optically-thin case, the efficiency is much lower. Local
absorption is neglible, and a flux of high-energy photons is
produced. These photons may interact with baryons by
photoionization, Compton scattering, and pair creation, and with
cosmic-radiation-background photons by photon-photon
scattering. After interaction, the energy is transferred to the
electrons, positrons, and ions which have stronger interactions
with other particles, or to photons with much smaller energy and
greater optical depth.  Here we make the following approximation:
Once a scattering happens at redshift $z$, the energy carried by
the photon at that redshift is completely deposited with some
efficiency. This seems to be a good approximation, since the electrons 
and lower-energy photons produced in the scattering event have
much greater optical depths.  The energy deposition rate is then 
\begin{eqnarray}
Q(z) = 4\pi \int dE && \left[ \chi_b n_b(z) \sigma_{b\gamma}(E) +
\chi_{\gamma} n_{\gamma}(z) 
\sigma_{\gamma\gamma}(E,z) \right] \nonumber \\
&& \times ~~ F(E,z)
\end{eqnarray}
where $\chi_b, \chi_{\gamma}$ are efficiencies for converting the
energy of that photon to ionization energy, and 
$\sigma_{b\gamma}(E)$ and $\sigma_{\gamma\gamma}(E,z)$ are the
cross sections for interacting with baryons and background photons,
respectively, the latter also depending on $z$ since the background
photon energy changes.
In the approximation of low optical depth (neglecting absorption),
the flux is given by
\begin{equation}
\label{eq:flux}
F(E,z) = \frac{c}{4\pi} \int_{z}^{\infty} \frac{dz'}{(1+z')H(z')}
\frac{J[\frac{1+z'}{1+z}E, z']}{(1+z')^3/(1+z)^3},
\end{equation} 
where $J(E,z)$ is the emissivity at $z$. For simplicity, let us
consider the case where the decay process directly produces photons with a
single energy $E_{\gamma} = x M_X c^2$. For example, if the decay
products are two photons, $x=1/2$. In other cases, the photon may not
have a single energy, but still it is expected the energy of the
photon is related to the mass of the decaying particle and has a
narrow range. Then $J(E,z)= N_{\gamma} M_X c^2 n_X(z) \Gamma_X \delta(E- x
M c^2)$, where $N_{\gamma}$ is the number of photon emitted in a
decay, and the flux is given by
\begin{equation}
\label{eq:fluxdelta}
F(E,z)= \frac{c}{4\pi} \left(\frac{E}{x M_X c^2}\right)^3
\frac{N_{\gamma} \Gamma_X n_X(z')}{H(z')}\left|_{\frac{1+z'}{1+z}=
\frac{x M_X c^2}{E}}\right. .
\end{equation}
Integrating over $E$, if the interactions with baryons dominate, then 
\begin{equation}
Q(z) \sim \frac{c n_b(z) \sigma_{\rm eff}}{H(z)} Q_0(z),
\end{equation}
where $Q_0(z)$ is the expression given in 
Eq.~(\ref{eq:Qinstant}) with $\chi_i = N_{\gamma} \chi_b$,
\begin{equation}
\sigma_{\rm eff} = \int \frac{dE}{x M c^2} \sigma_{b\gamma}(E)
\left(\frac{E}{x Mc^2}\right)^{3/2}.
\end{equation}
The efficiency is roughly suppressed by a factor of $\tau_S \sim
\frac{n_b(z) c \sigma}{H(z)}$, i.e. the optical depth for the
Hubble length.  If interaction with background radiation photons
dominate, the suppression factor is $\tau_S \sim n_{cmb}(z) c
\sigma(z)/H(z)$.  The redshift dependence is stronger due to the
additional factor of $1+z$ in the cross section.

\begin{figure}
\epsfig{file=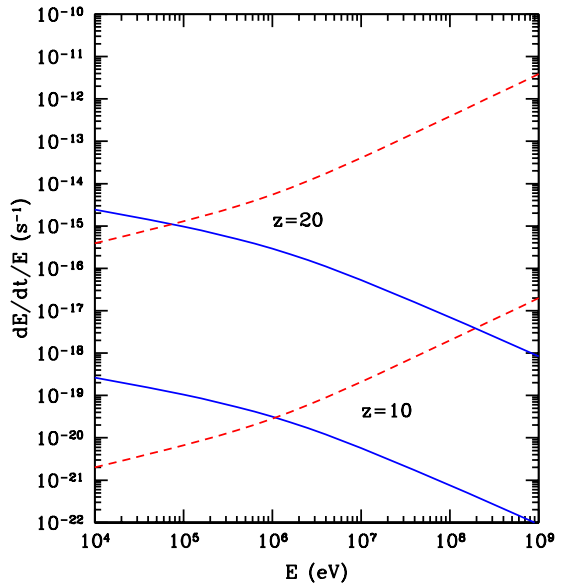,width=0.45\textwidth}
\caption{\label{eloss} Energy loss rate due to ionization (blue solid line) and
inverse Compton scattering (red dashed line) 
of an energetic electron. We plot for the cases of 
$z=10$ and  $z=20$.}
\end{figure}

\subsection{Electrons}

An electron can collide with and ionize atoms, or it can
inverse-Compton scatter CMB photons, a process that produces
an energetic photon.  Those photons will be absorbed again,
starting an electromagnetic shower. The energy loss $dE$ of an
electron per unit distance $dx$ by ionization in a neutral hydrogen gas
is given by \cite{Lang}
\begin{eqnarray}
-\frac{dE}{dx}= &&\frac{2\pi e^4 n_H}{m v^2} \left[\ln\frac{(mv^2\gamma^2
-T_m)}{2I^2}+ \frac{1}{\gamma^{2}}\right. \nonumber\\
&&\left. -\left(\frac{2}{\gamma}-\frac{1}{\gamma^{2}}\right)\ln 2  +\frac{1}{8}
\left(1-\frac{1}{\gamma}\right)^2\right], 
\end{eqnarray}
where
\begin{equation}
T_m = \frac{2 \gamma^2 m_H^2 m_e v^2}{m_e^2 +m_H^2 + 2\gamma m_e m_H} .
\end{equation}
The energy loss by inverse Compton scattering is given by 
\begin{equation}
-\frac{dE}{dt} = \frac{4}{3} \sigma_T c U_{CMB} \gamma^2.
\end{equation}
Other forms of energy loss are relatively unimportant for
reasonable values of parameters. For example,
synchrotron-radiation loss is
\begin{equation}
-\frac{1}{E}\frac{dE}{dt} = 1.05 \times 10^{-31} \frac{E}{\MeV}
\left(\frac{B}{\mu{\rm G_s}}\right)^2. 
\end{equation}
Since the energy-loss rate for inverse-Compton scattering is
proportional to $\gamma^2$, it dominates at high energy. 
At low energy, the ionization-loss rate is given approximately
by
\begin{equation}
-\frac{dE}{dx} \approx  2.54 \times 10^{-19} n_H (3\ln\gamma +
20.2)~ \eV/\cm.
\end{equation}
Since $n_H \propto (1+z)^3$ and $U_{CMB} \propto (1+z)^4$,
inverse-Compton scattering dominates the energy loss at
\begin{equation}
   z \gtrsim 20.8 \left(\frac{\Omega_b h^2}{0.022}\right)
   \left(\frac{2.726}{T_0}\right)^4 \gamma^{-2}.
\end{equation}

We plot the the energy-loss rate as a function of $E$ for $z=0$ and
$z=20$ in Fig.~\ref{eloss}. Generally speaking, if the electron has
energy greater than $\sim 100$ eV but smaller than $\sim$ MeV, the
energy-loss mechanisms are collisional ionization and excitation.
If the electron has energy greater than $\sim \MeV$, it loses
most of its energy by inverse-Compton scattering, producing UV
and X-ray photons.  If the electron energy is $E_e \lesssim$ GeV or
$E_e \gtrsim 50$ TeV, these photons are subsequently absorbed by
photoionization and excitation (or by pair production), and the
decay energy is thus transferred locally to the IGM.  However,
if the electron has an energy $1~\GeV \lesssim E_e \lesssim 50~\TeV$, the
up-scattered CMB photon has an energy in the transparency
window 10 keV--10 TeV.  Thus, if the injected electron has an
energy in this ``electron transparency window,'' the decay
energy will escape and appear in diffuse photon backgrounds, and
will not transfer most of its energy to the IGM.

For the case where the electron does heat the IGM, the partition
of the energy  among ionization, excitation, and heating was
investigated by Shull and Van Steenberg \cite{SvS85}.  They
found that when the gas is mostly neutral, about 1/3 of the
energy goes to ionization, about the same amount goes into
excitation, and the rest heats the IGM.  For a
fully ionized medium, almost all of the energy goes into heating
the gas. Therefore, we can approximate the fraction of energy
going into ionization as,
\begin{equation}
\label{eq:eff}
\chi_i \sim \chi_e \sim (1-x_e)/3, \quad \chi_h \sim (1+2 x_e)/3.         
\end{equation}
This approximation is crude but sufficiently accurate for our
purposes.

If the energy of the initial electron is still higher, $E \sim
1~ \MeV^2 /E_{CMB} \sim 10^{10} (1+z)^{-1}~\MeV$, it can scatter with
another photon or electron and produce an electron-positron
pair. The electron and positron then lose energy through
inverse-Compton scattering or ionization, the positron
eventually annihilates with another electron and produces $511
~\keV$ photons.  To summarize, roughly one-third the decay
energy goes locally into ionization, and the rest into heating
the gas, unless the electron is injected in the transparency
window GeV$\lesssim E \lesssim 50$ TeV, in which case most of the
energy is carried away by upscattered CMB photons.

\subsection{Other particles}

{\it Protons} are very penetrating particles and thus are not
effective in transferring decay energy to the IGM.

{\it Other particles.} Other decay products (e.g., muons,
tau leptons, heavy quarks, gauge, or Higgs bosons) will
generally produce showers of lower-energy particles.  This is a
complicated and model-dependent process.  In the absence of a
given well-motivated candidate that decays to these particles,
we neglect to carry out a detailed analysis.  Roughly speaking,
we expect typically 10\% of the decay to wind up in ionization
energy at the decay redshift, with a comparable amount going to
heating the gas, and the rest begin carried away by neutrinos or
as rest-mass energy of decay particles.

\section{Ionization History}

The decay of an unstable particle can affect both recombination
at $z \sim 1000$ and reionization at low $z$, or it may peak at
a middle redshift. However, since the first acoustic peak is not
significantly damped, recombination must be rapid, and completed
well before $z \sim 100$, when the Universe starts to become 
optically thin even at full ionization \cite{PSH00}.

In the presence of the decaying particle, the evolution of the
ionization fraction $x_e$ satisfies
\begin{equation}
\frac{d x_e}{d z} = \frac{1}{(1+z)H(z)} 
\left[ R_s(z) - I_s(z) - I_X(z)  \right],
\end{equation}
where $R_s$ is the standard recombination rate, $I_s$ the ionization
rate by standard sources, and $I_X$ the ionization rate due to
particle decay. This last term is related to the energy-deposition rate $Q$
introduced earlier: $I_X = Q(z)/n_b(z)/E_0$, where $E_0$ is the average
ionization energy per baryon.
At low redshift, the standard ionization sources are photons
from stars or active galactic nuclei (AGN), $I_s = I_*$, and the
standard recombination rate is
\begin{equation}
R_s = C_{\rm HII} \alpha_B(T) x_e^2 n_b(z) ,
\end{equation}
where $\alpha_B(T)$ is the case B recombination coefficient for
gas at temperature $T$ and density $n_b$. Here $C_{\rm HII}$ is
the clumping factor. We take $C_{\rm HII}=1$, appropriate for
$z\gtrsim20$ \cite{jordietal}.

The number density of decaying particles is proportional to
$(1+z)^3 e^{-\Gamma_X t}$, and its energy density
is simply the number density times the rest mass.  The
particle-decay ionization rate is
\begin{equation}
I_{Xi} = \epsilon_{Xi}(z) H, \quad {\rm with} \quad
\epsilon_{Xi}(z) = \chi_i(z) \frac{M_X}{E_0} \frac{n_X(z)}{n_b(z)}.
\end{equation}
To simplify the analysis, we neglect the effect of helium and
assume $m_b=m_H$ and $E_0=13.6$.  The partition of ionization
energy in hydrogen and helium depends on the nature of the
decaying particle.  The helium atom has a greater ionization
energy and also a greater photonionization cross section, so it
will probably take away more energy than hydrogen and produce
fewer electrons. However, it should not affect the order of
magnitude of our estimate.  
The ratio $m_b/E_0 \sim 7 \times 10^7$; thus even only a tiny
number of particle decays may supply enough energy to reionize
the Universe.

Since there are no stars present at recombination, CMB photons
are the main source of ionization. In this case, a recombination
to the ground state produces an ionizing photon which
immediately ionizes another atom and thus produces no net
change in the ionization fraction; only recombination to the $n\ge 2$
state produce net recombination. Assuming the number of photons
in the $2s$ state given by the thermal-equilibrium value, the
net recombination rate is
\begin{equation}
R_s -I_s = C \left[\alpha_B(T) x_e^2 n_b(z) - \beta_T (1-x_e)
e^{-E_{2s}/kT_M}\right],
\end{equation}
where $\beta_T$ is the photoionization coefficient, 
\begin{equation}
C=\frac{1+K \Lambda n_b (1-x_e)}{1+K (\Lambda+\beta) n_b (1-x_e)},
\end{equation}
and $\Lambda = 8.23\, {\rm s}^{-1}$ is the two-photon decay rate
of the 2s level. During this epoch, particle decay increases the
ionization rate not only by direct ionization from the ground state,
but also by contributing additional Lyman-alpha photons which 
boost the population at $n= 2$, increasing the rate of
photoionization by the CMB from these excited states,
\begin{equation}
I_X(z) = I_{Xi}(z) + I_{X\alpha}(z) ,
\end{equation}
where $I_{Xi}$ is the ionization rate given above, and $I_{X\alpha}$
the ionization rate due to additional Lyman alpha photons,
\begin{equation}
I_{X\alpha} = (C-1) \epsilon_{\alpha}(z) H, \quad \quad
\epsilon_{X\alpha}(z) = \chi_{\alpha}(z) \frac{M_X}{E_{\alpha}}
 \frac{n_X(z)}{n_b(z)} \frac{\Gamma_X}{H}. 
\end{equation}
Using $n_X=\Omega_X \rho_c/M_X$ and $n_b=\Omega_b \rho_c/m_b$, where 
$\Omega_X(z)$ is the fractional abundance of the decaying
particle at $z$, $M_X$ the mass of the decaying particle, and 
$m_b$ the mean baryon mass, we have 
\begin{equation}
I_{Xi}(z) = \chi_i(z) \frac{m_b}{E_0} f_X
\Gamma_X,\quad 
I_{X\alpha}(z) = \chi_{\alpha}(z) 
\frac{m_b}{E_{\alpha}} f_X(z)
\Gamma_X,
\end{equation} 
where $f_X = \Omega_X(z)/\Omega_b(z)$.
Written in this form, the ionization rate depends only on the
fractional abundance of the particle and the ionization
efficiency.  The gas-temperature evolution is given by
\begin{eqnarray}
(1+z)\frac{dT_b}{dz}&=&\frac{8\sigma_T a_R T_{CMB}^4}{3m_e
c H(z)}\frac{x_e}{1+f_{\rm He}+x_e} (T_b -T_{CMB})\nonumber \\
&& -\frac{2}{3
k_B H(z)} \frac{K(x)}{1+f_{\rm He}+x_e} +2 T_b,
\end{eqnarray}
where 
\begin{equation}
K(x)=\chi_h m_b \frac{\Omega_X(z)}{\Omega_b(z)} \Gamma_x.
\end{equation}
We use a modified version of the code {\tt RECFAST}
\cite{recfast} to calculate these rates and derive the
ionization history of the gas for a given decaying particle.

Once the ionization history is obtained, the CMB anisotropy can
be calculated by modifying a standard Boltzmann code.
We have used {\tt CAMB} \cite{camb} for our calculation. 
Except for the power-spectrum normalization and reionization
optical depth, we adopt the WMAP best-fit parameters for the
flat $\Lambda$CDM model with a power-law spectrum \cite{WMAP}; i.e. 
$\{\Omega_{m0} h^2,~ \Omega_{b0} h^2,~ h,~ n_s\}=\{0.14,~ 0.024,~
0.72,~ 0.99\}$.

\subsection{Decaying particle with long lifetime}

The decaying particle may or may not be a major component of the
dark matter. If $\Gamma \ll H_0$, and the primary decaying
particle is massive, then $\Omega_x(z)/\Omega_b(z)$ is
effectively constant, and the ionization history depends only on
the energy output $\xi=\chi_i f_x \Gamma_x$.  After recombination, the
ionization fraction induced by the decaying particles may be
estimated from the Saha equation as,\footnote{If there is no extra
ionization, the Saha equation does not necessarily describe the
ionization state as ionization reactions may have already frozen
out because of the paucity of free electrons.}
\begin{equation}
x_e = (\epsilon_i H/\alpha_B n_b)^{1/2} \propto (1+z)^{-3/2}.
\end{equation}
As the Universe expands, the physical density drops, the
recombination rate decreases, and the ionization fraction
increases, until at a certain point the Universe is fully
ionized, or ionization by decaying particles is exceeded by
stellar reionization. The contribution to the optical depth is
then
\begin{equation}
\tau = \int \frac{c dz}{H} \sigma_T x_e(z) n_{b0} (1+z)^2
\propto \ln (1+z).
\end{equation}

\begin{figure}
\epsfig{file=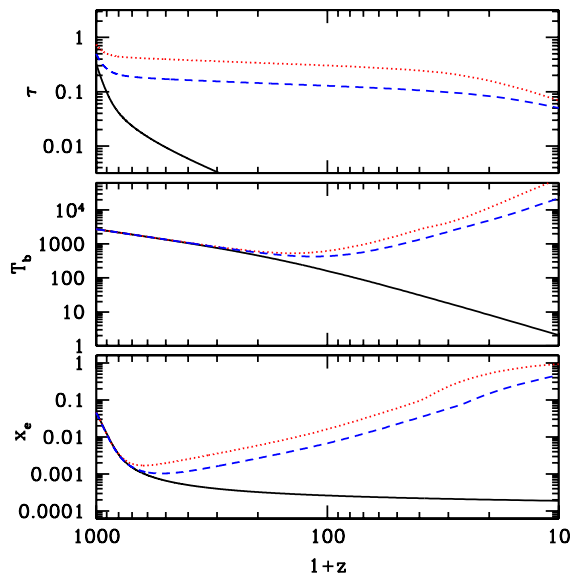,width=0.45\textwidth}
\caption{\label{Fig:reclong} The optical depth, IGM temperature,
and ionization fraction as a function of redshift for standard
recombination with no reionization (black solid line) and a
decaying-particle model with two-particle decay with $\Gamma_X
\ll H_0$ and $\xi \equiv \chi_i f_X \Gamma = 2.4\times 10^{-23}
~\persec, \tau=0.4 $  (red dotted line), and  $0.6
\times 10^{-23} ~\persec, \tau=0.17 $ (blue dash line).
}
\end{figure}

\begin{figure}
\epsfig{file=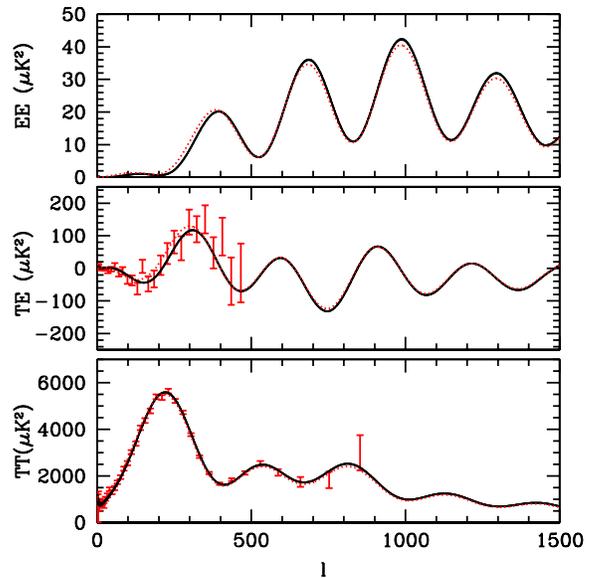,width=0.45\textwidth}
\caption{\label{Fig:cllonga} The CMB temperature and polarization
power spectrum $l(l+1)C_l/(2\pi)$ for decaying particles with
lifetimes greater than the age of the Universe. The data points
with error bars are
the binned data given by the WMAP team \cite{wmapdata}. The curves
are: no particle decay (black solid line), long-lived particle decay
with $\xi= 2.4\times 10^{-23} ~\persec$ (red dotted line) and 
$0.6 \times 10^{-23}~\persec $ (blue dash line). 
}
\end{figure}

\begin{figure}
\epsfig{file=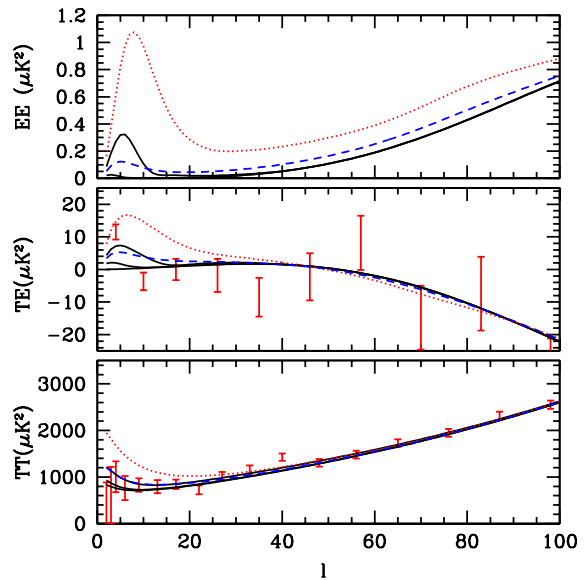,width=0.45\textwidth}
\caption{\label{Fig:cllong} Same as the previous Figure, but for 
$l<100$: $\xi= 2.4\times 10^{-23} ~\persec$ (red dotted line) and 
$0.6 \times 10^{-23}~\persec $ (blue dash line). We also plotted
three curves for the no-particle-decay case (black solid line) which
are almost indistinguishable except for the TE polarization;
from top to bottom they are $\tau=0.17$, step-function
reionization at $z<7$, and no reionization. 
}
\end{figure}

In Figs.~\ref{Fig:reclong} we plot the ionization history for
the case of a long-lived decaying particle ($\Gamma_X \ll H_0$)
and instant energy deposition, but with different energy output
$\xi$.  In this Figure, $\tau$ is the Thomson optical depth
between today and redshift $z$; $T_b$ is the gas temperature;
and $x_e$ is the ionization fraction.
At $z \sim 1000$ the ionization fraction drops rapidly due
to rapid recombination and to the decrease of the ionization rate by the CMB.
When particle decay starts to dominate the ionization rate
at $z \sim 600-800$, the ionization fraction starts to increase again
because now the ionization rate is constant while the recombination rate
drops. At $z \sim 100$, the ionization fraction can reach a few
percent, two orders of magnitude higher than in standard models. 
The optical depth increases slowly with $z$.

The temperature of the IGM also starts to increase at $z \sim 100$, and 
continues to climb as $z$ decreases. The reason for this is that as 
the neutral fraction decreases, with the efficiency assumed in 
Eq.~(\ref{eq:eff}), more and more energy is converted to heat at low
redshift, and also at lower redshift the baryons and 
CMB photons are kinetically decoupled. The difference in the 
CMB and gas temperature produces distortions to the CMB
blackbody spectrum, which are quantified by the Compton-$y$
parameter,
\begin{equation}
y= \sigma_T c \int  \frac{k_B (T_e - T_{CMB})}{m_e c^2}
\frac{x_e n_{b}(z) dz}{(1+z) H(z)}.
\end{equation}
However, for this and all other models studied in this paper, we found
this effect induces $y <10^{-8}$, well below the current
limit \cite{F96}.

How does the additional ionization by particle decay affect the CMB
anisotropy? As is well known (see \cite{TEreion} and references
therein), for the temperature anisotropy there is only a weak
effect on large  scale, but on small scales the temperature  
spectrum is damped by a factor of $e^{-2\tau}$. The division of large
and small scale is determined by the angular size of the horizon at the
reionization redshift.  In principle, the power-spectrum
normalization can also be determined by other measurements \cite{CFO03}.
However, other parameters also affect small-scale
anisotropy. To avoid re-fitting all the cosmological
parameters, we simply fix all other
parameters, and adjust the overall normalization to 
fit the WMAP TT and TE data.\footnote{We multiply the
unnormalized CMB power spectrum by a constant, which is then
adjusted  to minize $\chi^2$ with respect to the first year 
WMAP TT (up to $l=900$) and TE data (up to $l=512$)
\cite{wmapdata}.  We assume the errors are uncorrelated.}

So far we have considered 
only energy input from particle decay.  At low redshifts,
stars and quasars contribute a large part (if not all) of the
ionizing photons.  Since it is not the aim of this paper to
provide a detailed model of the star-formation history, we
simply illustrate the effects of particle decay on the CMB power
spectrum by using an ionizing flux due only to particle decay
for $z>z_*=7$; we then assume the Universe became suddenly
and permanently reionized by standard sources below that redshift.
We then calculate the CMB temperature and polarization
anisotropy for this ionization history. 

The renormalized CMB temperature and polarization power spectra
are plotted in Fig.~\ref{Fig:cllonga} (all $l$s)
and Fig.~\ref{Fig:cllong} (low $l$s).   
There is practically no difference in the $TT$ spectrum for the
different models at high $l$, although there are small
differences in the $TE$ and $EE$ spectrum. In Fig.~\ref{Fig:cllong}
we plot the low-$l$ results only; here the difference is more
apparent. Since the overall normalization is increased, the low-$l$ 
multipoles of the high-$\tau$ models are raised. Current
observations favor low power at large angular scales \cite{WMAP},
so these models are not favored. In the TE data, the spectrum peaks 
at $l \sim 10$, which is again in contrast to the data, which is low
at $l \sim 10$. The greatest difference, however, is in the EE power
spectrum, which should easily distinguish different models.

\subsection{Decaying particle with short lifetime}

\begin{figure}
\epsfig{file=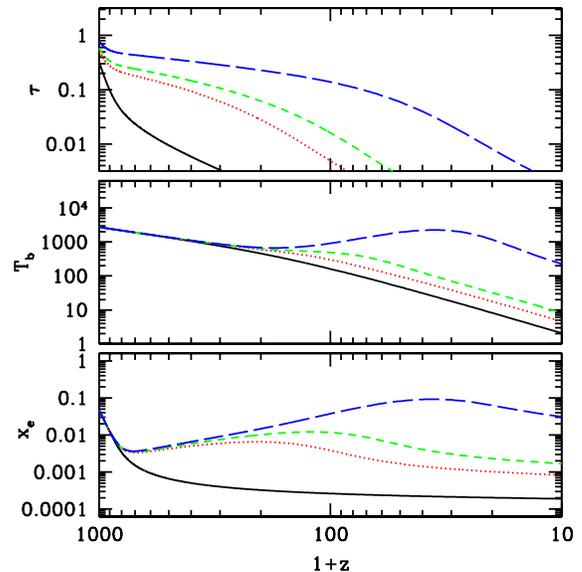,width=0.45\textwidth}
\caption{\label{Fig:recshort} The optical depth, IGM temperature, and
ionization fraction for the standard no-reionization model
(black solid line) and particle-decay-only models, all with
$\chi=0.3$, and $\Gamma_X=10^{-14}~\persec$, $f_X(z_{\rm
eq})=0.5 \times 10^{-8} $ (red dotted line);
$\Gamma_X= 0.5 \times 10^{-14}~\persec$,
$f_X(z_{\rm eq})=10^{-8} $ (green short dash line); 
and  $\Gamma_X=10^{-15}~\persec~ $, 
$f_X(z_{\rm eq})=5 \times 10^{-8} $ (blue long dash line). 
}
\end{figure}

\begin{figure}
\epsfig{file=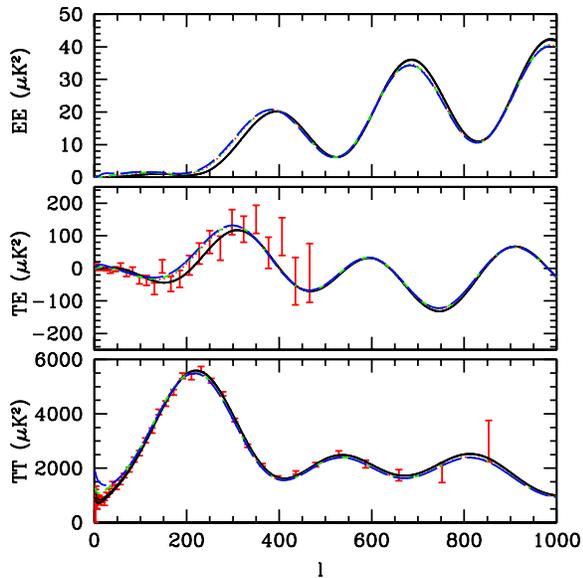,width=0.45\textwidth}
\caption{\label{Fig:rcllshort} The CMB temperature and polarization
power spectrum. Same models as the previous Figure.}
\end{figure}

\begin{figure}
\epsfig{file=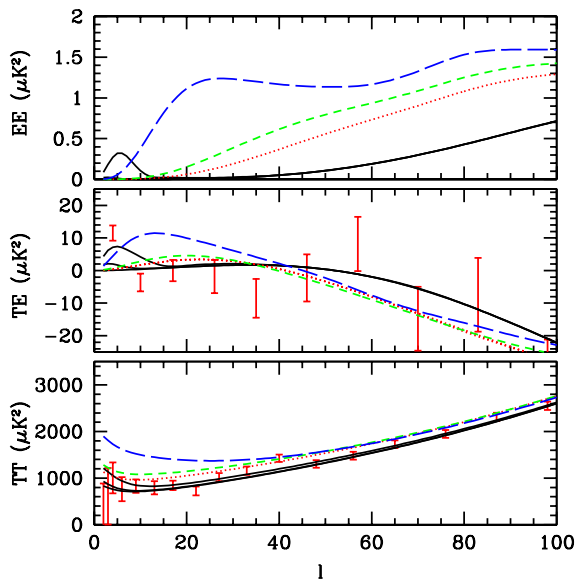,width=0.45\textwidth}
\caption{\label{Fig:rclshort} Low-$l$ CMB temperature and polarization
power spectrum. Same models as Fig.~\ref{Fig:rcllshort}. The three
black solid lines  (almost
indistinguishable except for the TE polarization) are, from top to
bottom,  for  $\tau=0.17$, step function reionization at
$z<7$, and no reionization. 
}
\end{figure}

If the primary particle has a lifetime less than the age of
the Universe, its density will change dramatically when the age
of the Universe is comparable to the lifetime. Moreover, its
density today will be small.  To be consistent with current
observations of the CMB, which are well fit by a matter density
comparable to that obtained from dynamical constraints in the
present-day Universe, the cosmological density of the decaying
particle must be small at the time of decoupling as well.  Thus,
$\Omega_X (z) \ll \Omega_m (z) \approx 1$. In this case, 
\begin{equation}
\frac{\Omega_X(z)}{\Omega_b(z)} \approx
\frac{\Omega_{X0}}{\Omega_{b0}}
e^{\Gamma_X (t_0 -t(z))} = \frac{n_{X{\rm eq}}}{n_{b{\rm eq}}} e^{-\Gamma_X (t(z)-t_{\rm eq})},
\end{equation}
where $n_{X{\rm eq}}, n_{b{\rm eq}}$
are the number density of decaying particle and
baryon at radiation matter equality, and the elapsed time is 
\begin{equation}
t(z) = \int_z^{\infty} \frac{dz}{(1+z) H(z)} \sim
\frac{2}{3}\frac{1}{H_0\Omega_0^{1/2}}
\frac{1}{(1+z)^{3/2}}.
\end{equation}

Unlike the long-decay-lifetime case, $\Gamma_X$ and $f_X$ must
now be treated as independent parameters.  We show a few
examples of the ionization history for short-lifetime decaying
particles in  Figs.~\ref{Fig:recshort} and \ref{Fig:rcllshort}.
The ionization rate is approximately
constant until $t(z) \sim \Gamma_x^{-1}$, after which it
decreases rapidly. In this scenario, the ionization fraction
increases slowly after recombination, just as in the long-lifetime
case. However, because the number density of the particle
decreases, the ionization fraction peaks broadly at a certain redshift
and then starts to decrease again.  The peak position depends on
the lifetime. The models plotted 
in Fig.~\ref{Fig:recshort} have $\Gamma_X^{-1} =  10^{14}$ s, $2
\times 10^{15}$ s, and $10^{15}$ s, 
which correspond to the age of the Universe at 
$z=300, 190$, and $65$ respectively.
Again, the ionization fraction can reach a few percent at $z\sim
100$ without jeopardizing the structure of the CMB acoustic
peaks. The temperature of the IGM
departs from the CMB temperature at redshifts of a few hundred in these cases,
but does not increase to a very high value
because of the decreasing energy available for heating. The
optical depth raises more sharply in this model, because there is more
variation in the free-electron density at high redshift. As a result,
the effect on the CMB is more apparent. We can see from
Figs.~\ref{Fig:rcllshort} and \ref{Fig:rclshort} that the CMB temperature 
as well as polarization peaks have different shapes, especially
apparent at high redshift. However, the TE correlation at low $l$
is less prominent for models with short lifetime, and even in
the model with relatively long lifetime, the peak is shifted to 
greater $l$, in strong disagreement with the WMAP
result. Inclusion of low-redshift reionization at $z=7$ results only in 
slight improvement. Based on this, particle decay with a short lifetime
does not appear to help solve the high TE optical depth as observed by
WMAP. 

\subsection{Additional redshift dependence}

\begin{figure}
\epsfig{file=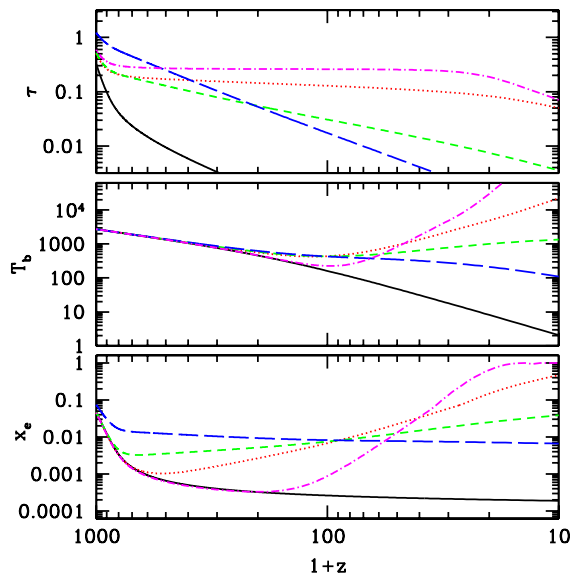,width=0.45\textwidth}
\caption{\label{Fig:longpowrec} 
The ionization fraction, temperature, and
optical depth for models with photons in the transparency window or 
have non-standard density evolution. The curves are
standard no reionization model (black solid line) and 
particle decay only models, all with $\tau_{S100}=10^{-3}$,
$\xi =0.6 \times 10^{-24}$, $n=0$ (red dotted line); 
$\xi =0.6 \times 10^{-24}$, $n=1.5$ (green short dash line); 
and $\xi =0.6 \times 10^{-25}$, $n=3$ (blue long dash line);
 $\xi =0.6 \times 10^{-25}$, $n= -3$ (magenta dash-dotted line). 
}
\end{figure}

\begin{figure}
\epsfig{file=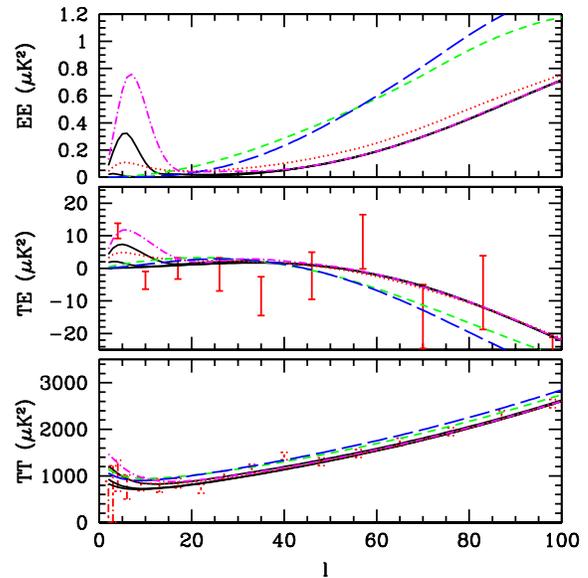,width=0.45\textwidth}
\caption{\label{Fig:longpow} The CMB temperature and polarization
power spectrum for the models shown in Fig.~\ref{Fig:longpowrec}.
}
\end{figure}

What if the decaying products are photons 
in the transparency window and deposit their energy differently?
As we discussed in Section II, the effect of photons in the 
transparency window in
the long-decay-lifetime case can be described with a suppression factor of 
$\tau_S \sim  n(z) \sigma_{\rm eff} c/H(z)$, which for baryons has a  
$(1+z)^{3/2}$ dependence. 

Additional dependence on the redshift may also raise  
if the density of the decaying particle does not vary 
as $(1+z)^3 e^{-\Gamma_X t}$. For example, if the decaying particle is 
relativistic, its energy density decreases as $(1+z)^4 e^{-\Gamma
t'}$, where $t'$ is the proper time of the moving particle. Also, 
the decaying particle could be continuously produced. 

We now consider these effects on the CMB by
multiplying $\xi$ with a factor $\tau_{S} (1+z)^n$, with $\tau_{S} \ll
1$. As an example, we consider models with 
$\tau_{S100}=10^{-3}$ at $1+z=100$, and $n=0, 1.5, 3$, and $-3$.
Obviously, at least for the $n=0$ case, the ionization induced by
particle decay would be uninterestingly small if we still use the 
same parameters as in \S III.A, since it is now suppressed by a factor
of $\tau$. To see the effect of $z$ dependence, we increase $\xi$
by a factor of 1000 for the $n=0$ and $1.5$ models, which cancels
the small $\tau_S$ value we assumed.
As it turns out, for the $n=3$ and $-3$
models this produces too large a deviation during the
recombination era which could easily be ruled out, so for the $n=3$ and $-3$
models
we increase $\xi$ by a factor of 100. The ionization fraction,
temperature, and optical depth are plotted in Fig.~\ref{Fig:longpowrec}.

For the cases with $n > 0$,
the additional redshift dependence $(1+z)^n$ makes the 
ionization energy redshift away. This means that if 
we adopt parameters which do not spoil standard recombination,
the effect on low redshift ionization must be very small. 
As shown in Fig.~\ref{Fig:longpow},
this does not help explain the WMAP result. 

{}From the above discussion, it seems that models with $n<0$ may produce 
results which are more consistent with the WMAP data. For example, 
the $n=-3$ model tends to have relatively large effect at lower $l$. If,
for example, the decaying particle is somehow 
associated with the dark energy which has a redshift dependence of 
$\rho \sim (1+z)^0$, then the $n=-3$ model might be realized.

Thus, we have investigated the ionization history and CMB for a variety of
models. It appears long-lifetime models may help produce the large TE
polarization at low $l$. Short life time models and models with additional
redshift dependence typically work less well, except for particles
or other energy sources whose density decrease slower than ordinary
matter.  

\section{Observational constraints}

So far we have illustrated the effect of particles decay
during the dark ages with several examples.  We now use
measurements of the CMB power spectra and diffuse
backgrounds to place constraints to the decay-particle parameter
space.  

\begin{figure}
\epsfig{file=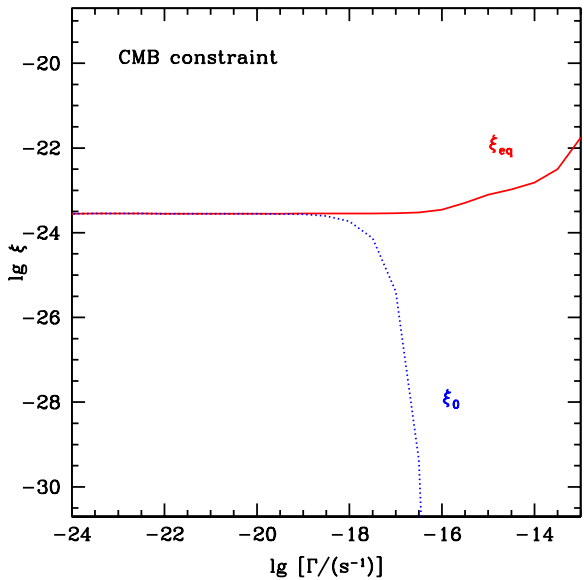,width=0.45\textwidth}
\caption{\label{fig:concmb} WMAP $1\sigma$ constraints on decaying
particles. Plotted are $\xi \equiv \chi f_{X} \Gamma_X$, where 
$f_X=\Omega_X/\Omega_b$. 
The red solid curve shows constraint on the value at 
matter radiation equality $\xi_{\rm eq}$, the blue dotted curve shows 
constraint on the value today $\xi_0$.
Note that the WMAP constraint applies if the injected photon or
electron energy does not fall in the transparency windows shown
in Fig. \ref{gcontour} and Section II.
}
\end{figure}

We first consider CMB constraints.  We suppose that all of the
decay energy is deposited instantly into the IGM, which then
reionizes the Universe and changes the CMB power spectrum as
described above.  For each set of model parameters $\Gamma_X$
and $\xi$ we calculate $\chi^2$ for the WMAP TT and TE
spectrum. We have 1403 TT and TE data points. The $\chi^2$
distribution for 1403 degrees of freedom has a a width of about
50, so the $1\sigma$ limit is defined by the line
$\chi^2_{min}+50$. The result is shown as the solid curve in
Fig.~\ref{fig:concmb}; regions above the curve are ruled out.
Note that these results do {\it not} apply if the particle
decays to photons or electrons in the transparency window, as in
these cases, the decay energy will not be deposited in the IGM
but will instead propagate freely and appear in the diffuse
radiation backgrounds.

In this case, the $\xi$-$\Gamma_X$ parameter space can be
constrained from measurements of diffuse backgrounds $z=0$.  The
flux of decay radiation is then given by Eq.~(\ref{eq:fluxdelta}).
with $z=0$. We then obtain bounds to the particle density and 
decay lifetime,
\begin{equation}
\label{eq:fluxobs}
N_{\gamma} \Gamma_X n_{X}(z) = \frac{4 \pi F(E)}{c} H(z),
\end{equation} 
where $1+z=x M_X c^2 /E$. 

We now apply this result to the X-ray and $\gamma$-ray backgrounds. 
The observed cosmic X-ray background (in units of $\cm^{-2} ~\persec
~\sr^{-1} $) can be modeled as \cite{kks03}
\begin{equation}
\label{eq:xobs}
F_X = \left\{ \begin{matrix}
8 \left(\frac{E}{\keV}\right)^{-0.4}, & 0.2~\keV < E < 25~\keV, \\
380 \left(\frac{E}{\keV}\right)^{-1.6}, & 25~\keV < E < 350~\keV, \\
2 \left(\frac{E}{\keV}\right)^{-0.7}, & 350~\keV < E < 2~\MeV.\\
\end{matrix} \right.
\end{equation}

The $\gamma$-ray background was measured by the 
the Energetic Gamma Ray Experiment Telescope (EGRET). At about 100
MeV, the flux is $1.45 \times 10^{-5} ~\cm^{-2} ~\persec ~\sr^{-1}$. A 
fit for the whole energy range of $30~ \MeV - 100~ \GeV$ was \cite{S98}
\begin{equation}
\label{eq:gobs}
F_{\gamma} = 1.73 \times 10^{-5} \left(\frac{E}{100~\MeV}\right)^{-2.11}
~\cm^{-2} ~\persec ~\sr^{-1}.
\end{equation}

More recent re-analysis found a greater galactic background, and 
therefore the extra-galactic background is lowered \cite{SMR03,KWL03}. 
Of course, distant quasars and stars also contribute to this 
background, so here we use the measured background Eq.~(\ref{eq:gobs})
as a conservative upper limit.

\begin{figure}
\epsfig{file=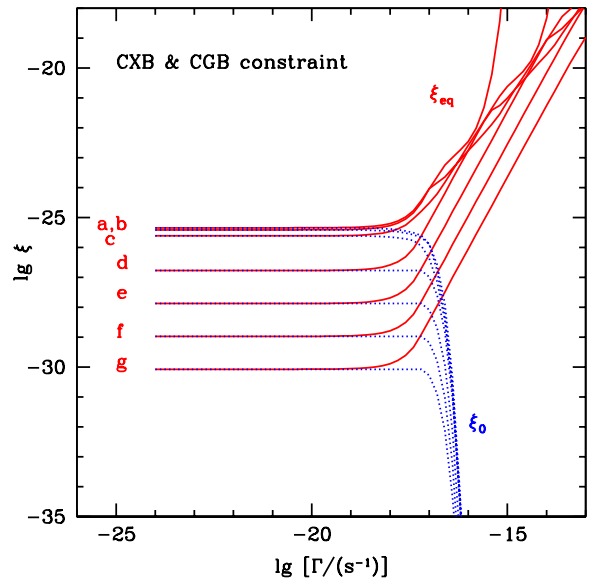,width=0.45\textwidth}
\caption{\label{fig:conxgb} Constraint of $\xi$ based on diffuse X-ray
and $\gamma$-ray background.  The red solid curve shows constraint on the value at 
matter radiation equality $\xi_{\rm eq}$, the blue dotted curve shows 
constraint on the value today $\xi_0$.
The curves are for photon energy (a) 100 keV, (b) 1 MeV, (c) 10 MeV, (d)
100 MeV, (e) 1 GeV, (f) 10 GeV, (g) 100 GeV. Note the x-ray and
$\gamma$-ray constraints do not apply for photon and electron
injection energies that fall outside the transparency windows.
}
\end{figure}

At each photon energy $E$, given the background flux $F(E)$, we can derive a 
bound on $n_{\rm eq}$ for each decay photon energy $x M_X c^2$ by 
applying Eq.~(\ref{eq:fluxobs}) to the flux (with the restriction that 
the maximum redshift to be less than $z \sim 1000$).
We run through the energy range of $1~\keV - 100~ \GeV$, at each
energy using the flux limits given in
Eqs.(\ref{eq:fluxobs}) and (\ref{eq:gobs}) to derive a bound, 
and look for the most restrictive constraint on $n_eq$ for all
energies. As expected, except for short-lived particles, the constraint comes
mainly from emission at $z=0$.
The results for decay photon in the transparency window 
$x M_X c^2 = 100 ~\keV, 1 ~\MeV$, $10 ~\MeV, 100 ~\MeV, 
1 ~\GeV, 10 ~\GeV, 100 ~\GeV$, are plotted as $\xi_{\rm eq}$
in Fig.~\ref{fig:conxgb}. On the same Figure we also plot the
corresponding values of today, $\xi_0$. 
In terms of the decaying particle at
radiation-matter equality, particles with short lifetimes are less
constrained than the ones with long lifetimes.
However, we do not expect any of
these short-lifetime particles to remain today, as shown in
Fig.~\ref{fig:conxgb}. 

Depending on the energy of the observed photon,
the diffuse X-ray and $\gamma$-ray background 
constraints are generally more stringent than the CMB constraint
except for short-lived particles. One must remember, however,  that
the CMB constraint applies only outside the transparency window,
whereas the $\gamma$-ray constraint applies in the transparency window.

\section{Conclusions}

In this paper we have investigated particle
decay during the dark ages. Such particle decay could induce partial
ionization of the Universe, and thus provide a potential alternative
to early star formation as an explanation for the WMAP TE
measurement.

\begin{figure}
\begin{center}
\epsfig{file=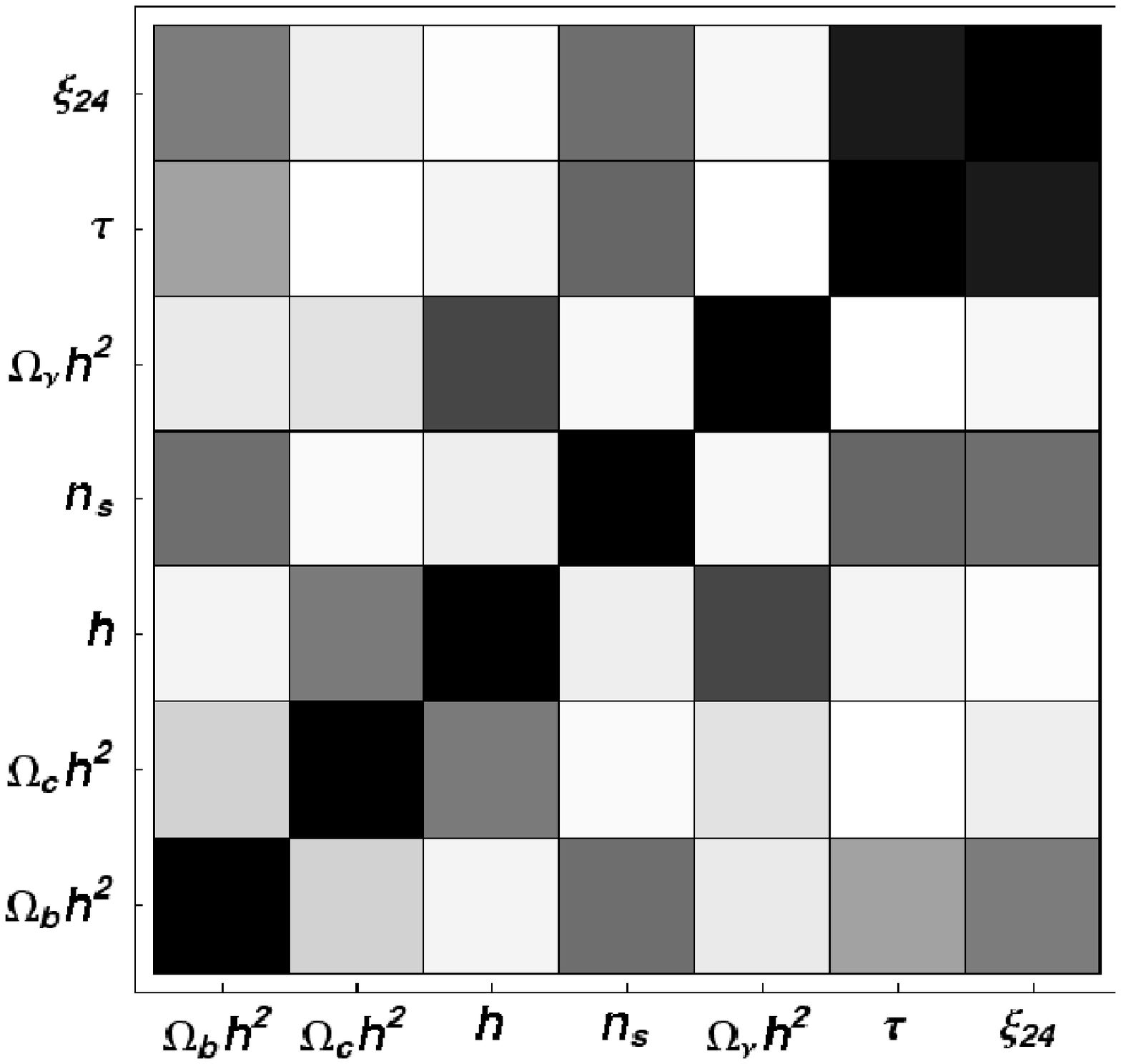,width=0.45\textwidth}\\
\epsfig{file=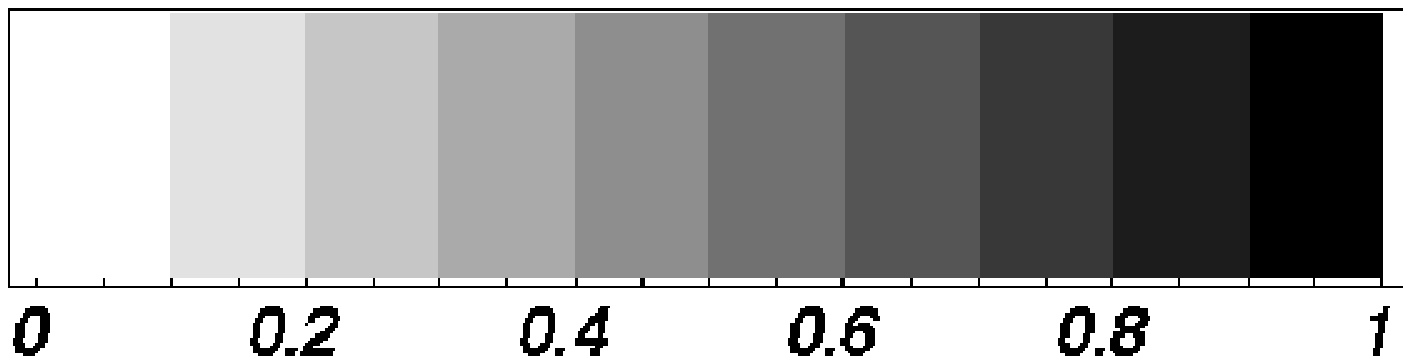,width=0.5\textwidth}
\end{center}
\caption{\label{fig:corr} The (squared) correlation 
matrix for cosmological parameters. The parameters are the physical
density of baryons $\Omega_b h^2$, cold dark matter $\Omega_c
h^2$, Hubble constant $h$, power index for primordial density
perturbation $n_s$, neutrino density $\Omega_{\nu}h^2$, Thomson
optical depth (see text) $\tau$, and the extra ionization parameter 
$\xi_{24} \equiv 10^{24} \xi$. We assume a flat Universe.}
\end{figure}

We considered how the decay energy is converted to ionization
energy. We conclude that in many cases, a shower of electrons and 
X-ray photons are produced, in which case 
a sizable fraction ($0.1-0.3$)
of the energy can be converted to ionization energy {\it in situ}, 
with comparable amount of energy going into heating the gas.
However, there are important exceptions. Photons in the energy range $100
~\keV-1~\TeV$ can escape, carrying with them most of the
energy. Electrons in the energy range $1~\GeV-50~\TeV$ lose most of
their energy by inverse-Compton scattering CMB photons into the above
energy range.
In these cases the ionization energy is deposited over a range of redshifts,
and the energy deposition is proportional to $\tau_S=n(z)\sigma_{\rm eff} c/H(z)$,
where $n(z)$ is the density of target particles, which can be 
baryons or the CMB photons. Because of the small value
of the optical depth $\tau_S$, the decay rate must be very large to affect
ionization. We can study  reionization in these models by
considering additional $1+z$ dependence.  In most cases, though,
the models will be ruled out, as seen in Fig. \ref{fig:conxgb}, by
diffuse backgrounds.

The extra energy input from particle decays can be parameterized by
the ionization energy input parameter 
$\xi =\chi f_X \Gamma_X$ ($\xi$ has the unit of $\persec$), 
where $\chi$ is the efficiency,
$f_X=\Omega_X/\Omega_b$, and $\Gamma_X$ the decay rate. 
If the lifetime of the decaying particle is longer than the age of the
Universe, the situation is particularly simple since the result
depends entirely on $\xi$. For short-lived particles, one must specify
both $\Gamma_X$ and $\xi$. We studied the ionization history and CMB 
temperature and polarization anisotropy for different cases. 
Although particle decays could partially ionize the Universe at high redshift
and produce a high optical depth, we found that in most cases they do not
reproduce the WMAP result very well. The TE polarization does not peak at
$l \sim 2$ but at $l \sim 10$, for example. We should pointed out
however, this is not a unique problem with particle-decay induced reionization,
but is also seen in other models with extended partial reionization history. 
We also found that the EE spectrum is
a sensitive probe to the ionization history. Furthermore, 
if reionization occurs at high redshift,
there is a change in the shape and position of the acoustic peaks.
The ionization history is affected if the extra energy input has 
additional dependence on the redshift. Typically, for an additional
redshift dependence of $(1+z)^n$ with $n>0$, the fit to CMB data is
not improved, because the extra energy input at early times will
spoil recombination.  Models with $n<0$ may be helpful, but
some exotic mechanism is needed for generating such a redshift
dependence.

\begin{figure}
\begin{center}
\epsfig{file=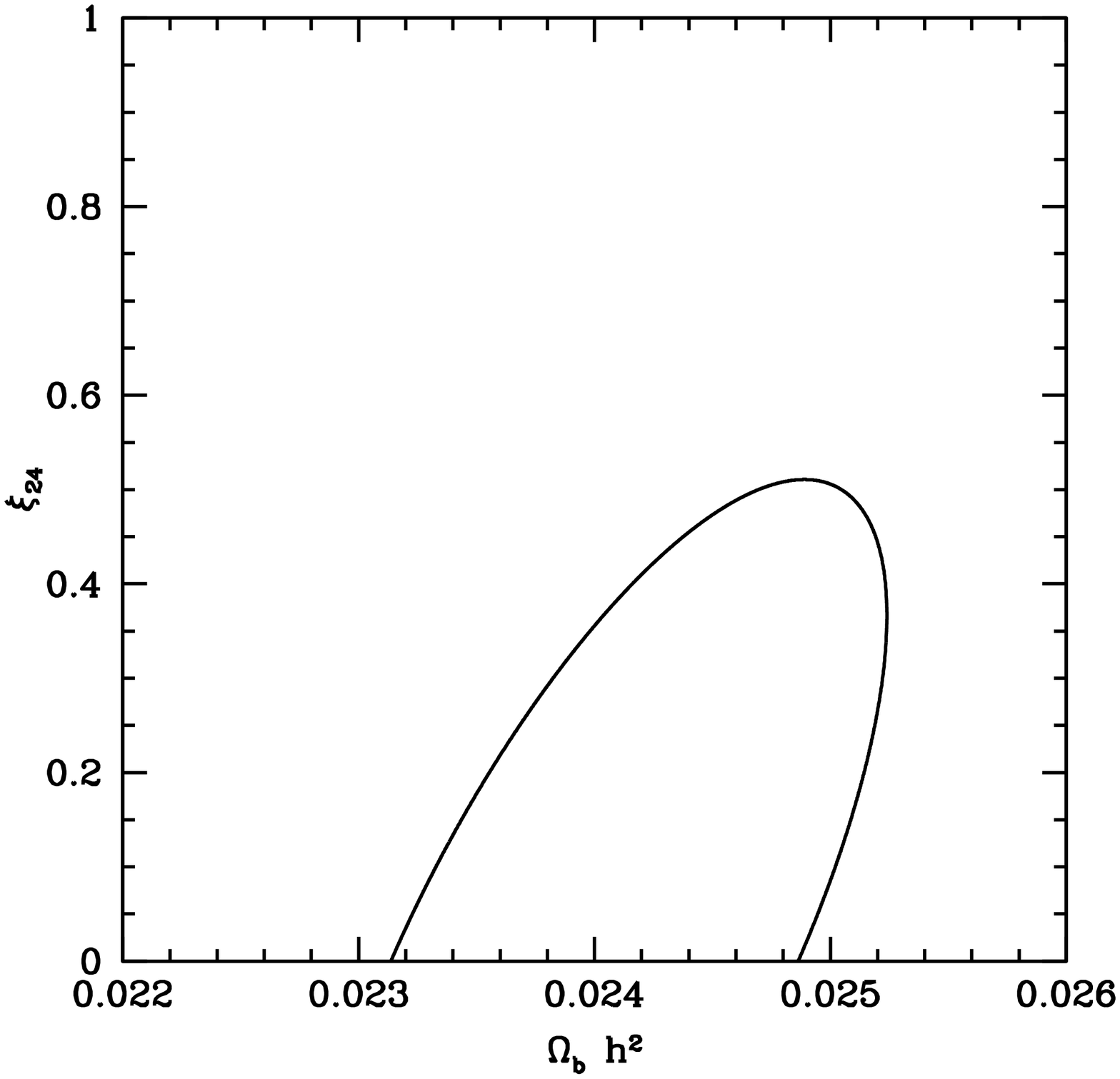,width=0.4\textwidth}\\
\epsfig{file=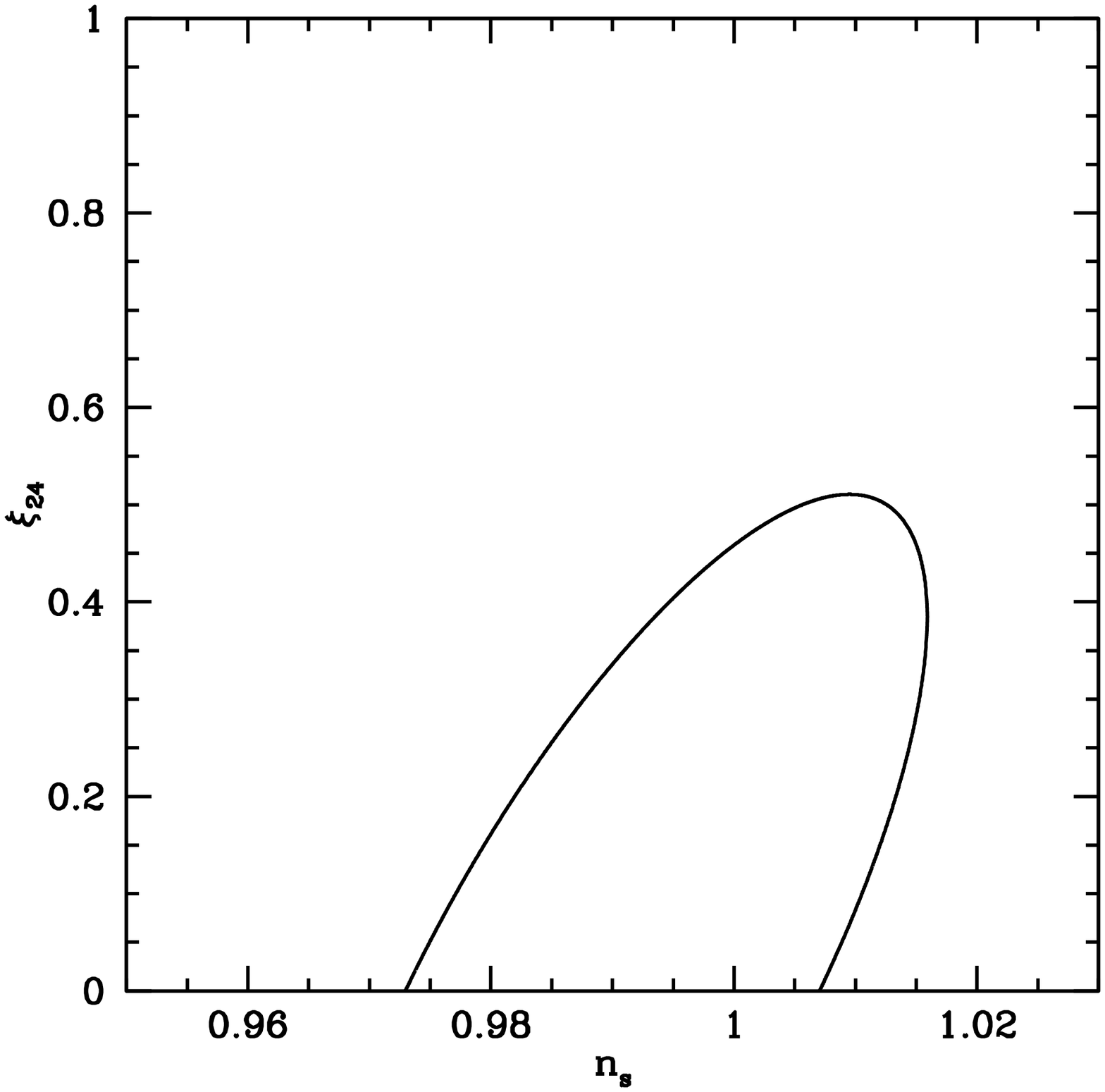,width=0.4\textwidth}
\end{center}
\caption{\label{fig:ellip} The error ellipses for $\Omega_b h^2$ vs. $\xi_{24}$
and $n_s$ and $\xi_{24}$. 
$\xi$ is positive from its physical interpretation. }
\end{figure}

We have obtained constraints on particle decays during the dark ages
using the WMAP data as shown in Fig.~\ref{fig:concmb}.
We found $\xi < 10^{-24}~{\rm s}^{-1}$ for the long-lifetime
case, and a slightly weaker bound for the short-lifetime
case. However, the short-lived particles decay at high redshift
and we do not expect to see any left today. We also obtained constraints on the
decaying particle from the observed diffuse X-ray and $\gamma$-ray backgrounds.
This constraint is generally more stringent than the CMB constraint, but it
actually applies to a different situation; i.e. the decay products
are mainly photons in the energy range of the transparency window,
where they can propagate freely across the Universe and
contribute very little energy to ionization. 

The extra energy input also heats up the IGM during the dark
age, and the temperature can rise to $10^{3-4}$ K.
Inverse-Compton scattering of free electrons can
induce distortion in the CMB blackbody 
spectrum, but the effect is unobservably small ($y < 10^{-8}$).

If the dark-matter particle can decay, it may affect the estimation of 
cosmological parameters. To see how each parameter is affected, we can
calculate the correlation matrix, which is
related to the covariance matrix  \cite{Jungman} by
\begin{equation}
\mathbf{r}_{ij} = \mathbf{C}_{ij}/\sqrt{\mathbf{C}_{ii} \mathbf{C}_{jj}},
\end{equation}
where the covariance matrix is given by the inverse of the Fisher
matrix $\mathbf{C} = \mathbf{F}^{-1}$, with 
\begin{equation}
\mathbf{F}_{ij} = \sum_l \left[\frac{1}{\sigma_{C^{TT}_l}^2} 
\frac{\partial C^{TT}_l}{\partial \theta_i} 
\frac{\partial C^{TT}_l}{\partial \theta_j} +
\frac{1}{\sigma_{C^{TE}_l}^2} 
\frac{\partial C^{TE}_l}{\partial \theta_i} 
\frac{\partial C^{TE}_l}{\partial \theta_j}\right],
\end{equation}
and ${\theta_i}$ are the cosmological parameters to be
estimated. We plot $r_{ij}^2$ in Fig.~\ref{fig:corr}.

In making Fig.~\ref{fig:corr}, we have taken a fiducial model with the 
WMAP best fits with the exception $\tau = 0.037$ which corresponds to
a sudden reionization with $z_{rei}=6.0$. Choosing different fiducial
models may affect the error estimates slightly. In addition to the standard
parameters (physical density of baryons $\Omega_b h^2$, 
cold dark matter $\Omega_c
h^2$, Hubble constant $h$, power index for primodrial density
perturbation $n_s$, neutrino density $\Omega_{\nu}h^2$, Thomson
optical depth $\tau$), we have added the extra ionization input energy
$\xi_{24} \equiv 10^{24}\xi$ in the long-lived decaying-particle 
case. We will not consider the short-live case since 
it is much more model dependent.
As expected, $\xi$ correlates strongly with the
Thomson optical depth due to low-redshift stellar light; it will
thus be difficult to distinguish them from CMB observations alone.
Both $\tau$ and $\xi$ correlates strongly with 
baryon density $\Omega_b h^2$ and the primordial spectral index
$n_s$. If a decaying particle exists but is neglected in the fit, then
results for the values of other cosmological parameters may be
biased, and the error bars may be underestimated. In 
Fig.~\ref{fig:ellip} we plot error ellipses for 
$\Omega_b h^2-\xi$ and $ n_s-\xi$ after marginalizing over the other
parameters.

There are other ways that particle decays during the cosmic dark
ages could play a role in cosmology.  Decays might affect 
the recombination process \cite{BMS03}.  Particle decays could
produce a surfeit of free electrons after recombination; these
extra electrons could then facilitate the formation of
${\rm H}_2$ molecules and thus potentially enhance the  
star-formation rate. On the other hand, particle decay may also heat
up the gas, thus increase the Jeans mass of the primordial gas and
suppress early star formation. The final outcome requires detailed
investigation which is beyond the scope of the present paper. 
Finally, if the contribution of the particle
to reionization is significant, it may not require formation of 
structure at high redshift, thus eliminating one objection to
warm dark-matter models.

With future experiments, we expect to obtain more precise
information on the ionization history than we have now, and
should a decaying particle with $t > 10^{13}$ sec exist, we may
discover it through indirect observations such as those
discussed here.

\acknowledgments
We thank R. Sunyaev, E. Scannapieco, S.-P. Oh, and M. Kaplinghat
for helpful comments and discussions. The {\tt CAMB} code we used in
this research was developed by A. Lewis and A. Challinor.
X.C. is supported by NSF
grant PHY99-07949, and M.K. is supported by NASA grant NAG
5-9821 and DOE grant DE-FG 03-92-ER40701.

\newcommand\AL[3]{~Astron. Lett.{\bf ~#1}, #2~ (#3)}
\newcommand\AP[3]{~Astropart. Phys.{\bf ~#1}, #2~ (#3)}
\newcommand\AJ[3]{~Astron. J.{\bf ~#1}, #2~(#3)}
\newcommand\APJ[3]{~Astrophys. J.{\bf ~#1}, #2~ (#3)}
\newcommand\APJL[3]{~Astrophys. J. Lett. {\bf ~#1}, L#2~(#3)}
\newcommand\APJS[3]{~Astrophys. J. Suppl. Ser.{\bf ~#1}, #2~(#3)}
\newcommand\MNRAS[3]{~Mon. Not. R. Astron. Soc.{\bf ~#1}, #2~(#3)}
\newcommand\MNRASL[3]{~Mon. Not. R. Astron. Soc.{\bf ~#1}, L#2~(#3)}
\newcommand\NPB[3]{~Nucl. Phys. B{\bf ~#1}, #2~(#3)}
\newcommand\PLB[3]{~Phys. Lett. B{\bf ~#1}, #2~(#3)}
\newcommand\PRL[3]{~Phys. Rev. Lett.{\bf ~#1}, #2~(#3)}
\newcommand\PRD[3]{~Phys. Rev. D{\bf ~#1}, #2~(#3)}
\newcommand\SJNP[3]{~Sov. J. Nucl. Phys.{\bf ~#1}, #2~(#3)}
\newcommand\ZPC[3]{~Z. Phys. C{\bf ~#1}, #2~(#3)}

\end{document}